\documentclass{aa}
\usepackage{graphicx}
\newcommand{\kmprs}  {\mbox{\rm km\,s$^{-1}$}}

\newcommand{\HI} {\ion{H}{i}}
\newcommand{\FeII} {\ion{Fe}{ii}}
\newcommand{\ZnII} {\ion{Zn}{ii}}
\newcommand{\CrII} {\ion{Cr}{ii}}
\newcommand{\SiII} {\ion{Si}{ii}}

\newcommand{\MgII} {\ion{Mg}{ii}}

\newcommand{\MgI} {\ion{Mg}{i}}
\newcommand{\lya} {Ly$\alpha$}
\newcommand{\EBV}{\hbox{$E(B\!-\!V)$}}
\def\ltsima{$\; \buildrel < \over \sim \;$}
\def\simlt{\lower.5ex\hbox{\ltsima}}
\def\gtsima{$\; \buildrel > \over \sim \;$}
\def\simgt{\lower.5ex\hbox{\gtsima}}
\begin{document}

\title{Zn and Cr abundances in damped Lyman alpha systems from the CORALS survey
\thanks{This work is based in part on observations collected at 
the European Southern Observatory, Chile (ESO Nos. 69.A-0051A \&\ 71.A-0067A) 
and at the W.M. Keck Observatory, which is operated as a scientific partnership 
among the California Institute of Technology, the University of California and 
the National Aeronautics and Space Administration. The Observatory was made
possible by the generous financial support of the W.M. Keck Foundation.}}

\titlerunning{Zn and Cr in CORALS DLAs}


\author{Chris~J.~Akerman \inst{1} 
\and Sara~L.~Ellison \inst{2}
\and Max~Pettini \inst{1} 
\and Charles~C.~Steidel \inst{3}}


\institute{
Institute of Astronomy, University of Cambridge, Madingley Road, Cambridge, CB3 0HA, UK.
\email{cja@ast.cam.ac.uk, pettini@ast.cam.ac.uk}
\and University of Victoria, Dept. Physics \& Astronomy, Elliot Building, 3800 Finnerty Road
Victoria, V8P 1A1, British Columbia, Canada.
\email{sarae@uvic.ca}
\and Palomar Observatory, California Institute of Technology, MS 105-24, Pasadena, CA 91125, USA
\email{ccs@astro.caltech.edu}
}

\date{Received ..... / Accepted ......}

\abstract
{
We present metal abundances in 15 damped \lya\  
systems (DLAs) from the Complete Optical and Radio 
Absorption Line System (CORALS) survey,  
designed to be free from any biasing effects due
to extinction of QSOs by dust in intervening absorbers.  
It has long been suggested that 
such biasing may explain differences in metallicity between
damped \lya\ systems and coeval luminous galaxies, and
between model predictions and observations.
We use our measured zinc and chromium abundances 
(combined with those for five more CORALS DLAs from the 
literature, giving us a very nearly complete sample)
to test 
whether the metallicity and degree of dust depletion in 
CORALS DLAs are significantly different from those
of existing, larger, samples of DLAs drawn
from magnitude limited, optical surveys.  
We find that the column density weighted metallicity of CORALS DLAs, 
${\rm [} \langle{\rm (Zn/H)}_{\rm DLA}\rangle {\rm ]} = -0.88 \pm 0.21$
in the redshift interval $1.86 < z_{\rm abs} < 3.45$,
is only marginally higher than that of a control sample
from the recent compilation by Kulkarni et al.,
${\rm [} \langle{\rm (Zn/H)}_{\rm DLA}\rangle {\rm ]} = -1.09 \pm 0.10$.
With the present limited statistics this difference is not
highly significant. 
Furthermore, we find no evidence for increased dust depletions
in CORALS DLAs---their [Cr/Zn] ratios conform to 
the known trend of increasing depletion (decreasing [Cr/Zn])
with increasing metallicity, and we have encountered no cases where Cr is 
as depleted as in local cold interstellar clouds. 
These results, when combined with the earlier findings
of the CORALS survey reported by Ellison et al. in 2001, 
make it difficult to invoke a dust-induced bias to explain
the generally low level of chemical evolution exhibited by
most DLAs.  Rather, they indicate that large 
scale optical QSO surveys give a fair census of the population of
high redshift absorbers.

}

\maketitle

\section{Introduction}

Spectroscopic studies of absorption line systems along quasar 
sightlines are an important source of information regarding the 
chemical evolution history of neutral gas in the universe.  Damped 
\lya\ systems (DLAs, defined to have N(\HI)\,$\ge~2~\rm{x}~10^{20}$ 
atoms~cm$^{-2}$; Wolfe et al. \cite{wolfe86}), which make up the high 
column density end of 
the distribution of absorption line systems, are a particularly 
interesting subset of absorbers to study.  They dominate the 
neutral hydrogen content of the universe available for star 
formation up to $z\,\ge~4$ (e.g. Storrie-Lombardi \&\ 
Wolfe \cite{storrie00}; Peroux et al. \cite{peroux03}) 
and are thought to be the progenitors of the disk galaxies we see today.
DLAs have therefore been used as a tool for probing the evolution of
galaxies, especially at high redshift where direct measurements are
more difficult.  
Wolfe et al. (\cite{wolfe95}) found that the comoving mass density 
in neutral hydrogen in DLAs, $\rm{\Omega_{DLA}}$, is similar to the 
mass density in stars at redshift $z\,= 0$, and therefore proposed
that its redshift evolution, d$\rm{\Omega_{DLA}}$/d$z$, could be 
taken as a global measure of the rate at which gas in galaxies is 
converted into stars.

The metallicity evolution of galaxies has also been probed by 
chemical abundance measurements in DLAs (e.g. 
Pettini et al. \cite{pettini97b}, \cite{pettini99};
Prochaska \&\ Wolfe \cite{prochaska99}, \cite{prochaska00}), which 
provide the most detailed information on the chemical composition 
of high-$z$ galaxies.
If indeed DLAs are representative of the galaxy population at a given
redshift, then one would expect that the evolution of their neutral 
hydrogen fraction and metallicity would track that
of the general galaxy population as a whole.  However, no strong 
evolution of either $\rm{\Omega_{DLA}}$ or metallicity ($\rm{Z_{DLA}}$) 
has been seen in studies from $z\,\sim~3.5$ down to $z\,\sim~0.5$ 
(e.g. Rao \&\ Turnshek \cite{rao00}; Ryan-Weber et al. \cite{ryan-weber03}; 
Kulkarni et al. \cite{kulkarni05}).

Furthermore, it is now well established that 
DLAs are generally at the low end of the metallicity
distribution of galaxies at redshifts $z = 2 - 3$
(see, for example, Figure 32 of Pettini \cite{pettini04}).
The typical DLA metallicity at this epoch
is only $Z_{\rm DLA} \simeq -1.2$, or $\approx 1/15$ of solar
(Pettini et al. \cite{pettini99}; Kulkarni et al. \cite{kulkarni05}),
while near-solar metallicities are common for 
luminous galaxies detected directly in their rest-frame
ultraviolet, optical, and far-infrared light 
(e.g. Pettini et al. \cite{pettini02}; Shapley et al. \cite{shapley04}; 
de Mello et al. \cite{de_mello04}; Swinbank et al. \cite{swinbank05}).

This difference could have a number of causes.
Selecting galaxies through 
\HI\ absorption may preferentially pick out
chemically unevolved systems, either because 
galaxies with generally low rates of star formation
dominate the cross-section (Mo, Mao \& White \cite{mo98}),
or because the \HI\ cross-section is largest 
during a stage prior to the onset of star formation.
However, direct imaging of DLA galaxies shows a 
varied population of hosts which span a range of
luminosities and morphological types
(Boissier, P\'{e}roux, \& Pettini \cite{boissier03};
Rao et al. \cite{rao03}; Chen \& Lanzetta \cite{chen03}; 
Weatherley et al. \cite{weatherley05}---see also
Ryan-Weber et al. \cite{ryan-weber03}).
Abundance gradients may contribute to the difference
in metallicity between the outer regions---which
offer the larger cross-section for absorption---and
the inner regions of galaxies where star formation 
activity is more prominent (Pettini et al. \cite{pettini94};
Chen, Kennicutt, \& Rauch \cite{chen05}; Christensen et al. \cite{christensen05}; 
Ellison, Kewley, \& Mall\'{e}n-Ornelas \cite{ellison05}). 
Quantitatively, however, the magnitude of such gradients 
has recently been questioned in nearby spirals 
(Bresolin, Garnett, \& Kennicutt \cite{bresolin04}),
and remains unknown at high redshifts.

A third possibility is that damped \lya\ systems
which are both metal- and gas-rich do exist, but 
are systematically underrepresented in current
samples drawn from magnitude-limited QSO surveys.
The hypothesis is that even moderate amounts of dust 
associated with intervening galaxies 
may be sufficient to preferentially exclude 
reddened QSOs from optical surveys,
and that the statistics of DLAs would accordingly be 
skewed by such bias against dusty absorbers.
It is this third possibility which we
address in the present paper.

The idea of dust obscuration of QSOs has a long history
(e.g. Heisler \& Ostriker \cite{heisler88}) and received
observational support by the work of 
Pei, Fall, \& Bechtold (\cite{pei91})
who found the spectra of QSOs with DLAs in the 
redshift range $1.77 \leq z_{\rm abs} \leq 2.80$ to have statistically
steeper continuum slopes than those of a control sample. 
On the basis of these results,
Fall \& Pei (\cite{fall93}) proposed that between 10\% and 70\% 
of bright QSOs at $z = 3$ may have been missing from
optical samples.
This claim is now tempered by the recent re-analysis
by Murphy \& Liske (\cite{murphy04}) based on the much larger compilation
of QSO spectra made available by the Sloan Digital Sky Survey
(SDSS---Stoughton et al. \cite{stoughton02}). 
From the comparison of 70 QSOs lying behind DLAs 
at $2.0 < z_{\rm abs} < 4.0$) with
a control sample which is one order of magnitude 
larger, Murphy \& Liske concluded that the difference 
in the continuum slopes $\alpha$ between the two sets of 
spectra is 
only $\Delta \alpha = - 0.04 \pm 0.05$
(using the usual definition of the spectral index,
whereby the QSO continuum flux is a power law of the 
form $f_{\nu} \propto \nu^{\alpha}$),
corresponding to a limit
on the colour excess due to SMC-like dust-reddening of $\EBV\!<\!0.02{\rm
\,mag}$ ($3\,\sigma$).
This value is significantly
lower than $\Delta \alpha = - 0.38 \pm 0.13$
reported by Pei et al. (\cite{pei91}).
Similarly, Ellison, Hall \&\ Lira (\cite{ellison05a}) find  
$\EBV\!<\!0.05{\rm\,mag}$ ($3\,\sigma$) from a study of the 
optical to infrared colours of a subsample of CORALS QSOs 
with DLAs in the range $1.8 < z_{\rm abs} < 3.5$.

However, the SDSS results still do not preclude 
the possibility that highly obscured QSOs may by missing,
or underrepresented in optical samples
(since SDSS QSOs are optically selected).
Theoretically, such selection effects have been 
appealed to in order to reconcile the predictions of 
hydrodynamic simulations (Cen et al. \cite{cen03};
Churches, Nelson \& Edmunds \cite{churches04}; 
Nagamine, Springel, \& Hernquist \cite{nagamine04})
and galactic chemical evolution models 
(Prantzos \& Boissier \cite{prantzos00}) with the observations.
Observationally, their importance is suggested by
the apparent anti-correlation between neutral hydrogen
column density $N$(\HI) and metallicity $Z_{\rm DLA}$
first pointed out by Boiss\'{e} et al. (\cite{boisse98}).

It is to assess quantitatively the importance
of dust-induced bias for the statistics of DLAs that
the Complete Optical and Radio Absorption Line System
(CORALS) survey was originally conceived. 
As the name implies, this programme aims at measuring
the properties of DLAs in a complete sample of QSOs
selected at radio wavelengths, where dust obscuration
is not expected to be an issue.
In the first stage of the project, Ellison et al. (\cite{ellison01})
identified a sample of 22 DLAs from intermediate dispersion
spectroscopy of all the QSOs (66) with emission redshift
$z_{\rm em} \geq 2.2$ in the Parkes quarter-Jansky sample
of flat-spectrum radio sources 
(Jackson et al. \cite{jackson02}; Hook et al. \cite{hook03}).
The optical spectra were of sufficient quality 
to measure $N$(\HI) in the 22 DLAs, enabling 
Ellison et al. (\cite{ellison01}) to determine both the 
number density of DLAs per unit redshift, $n(z)$,
and the corresponding comoving mass density 
of neutral gas, $\Omega_{\rm DLA}$.
The values found, $n(z) = 0.31^{+0.09}_{-0.08}$
and $\log \Omega_{\rm DLA} = -2.59^{+0.17}_{-0.24}$
at a mean redshift $\mathrm{<}z\mathrm{>} = 2.37$,
are higher than the corresponding quantities
previously determined from optically selected QSO
samples, but only marginally so.
In particular, 
the CORALS survey did not uncover a 
population of high column density 
($\rm{N(\HI)\,>~10^{21}cm^{-2}}$) DLAs 
in front of faint QSOs. Within the limitations
imposed by the small size of their sample,
Ellison et al. (\cite{ellison01}) concluded that selection
effects due to intervening dust may at most account
for an underestimate by a factor of $\sim 2$ in
$\Omega_{\rm DLA}$. 

The \HI\ results alone, however, do not tell us
about the metal and dust content of CORALS DLAs 
and whether they are higher, on average, than
those of the optically selected DLAs which have
been studied extensively over the last fifteen 
years. These are the questions which we explore
in the present work. 

Specifically, we have conducted a follow-up
programme of high resolution spectroscopy of the 
CORALS DLAs aimed at measuring in particular
the abundances of zinc and chromium. 
Meyer, Welty \&\ York (\cite{meyer89}) and 
Pettini, Boksenberg \&\ Hunstead (\cite{pettini89}, \cite{pettini90}) 
first drew attention to
the diagnostic value of these two elements.
Both are iron-peak elements, whose
abundances track that of Fe to within 
$\pm~0.1 - 0.2~\rm{dex}$ in Galactic 
stars of metallicities
from solar to about 1/100 of solar
(Chen, Nissen \&\ Zhao \cite{chen04}; 
Cayrel et al. \cite{cayrel04} and references therein).
In the interstellar medium of the Milky Way,
on the other hand, Zn is one of the few elements
which show little affinity for dust grains, unlike
Cr which is usually highly depleted (Savage \& Sembach \cite{savage96}).
In combination, therefore, these two elements
can be used to obtain approximate measures
of the overall degree of metal enrichment,
via the [Zn/H] ratio, and the fraction of refractory
elements locked up in solid form, via the [Cr/Zn]
ratio.\footnote{We use the conventional notation whereby
[Zn/H]$ = \log{\rm (Zn/H)} - \log {\rm (Zn/H)_{\odot}}$.} 
Both elements have absorption lines
of their dominant ionisation stages in \HI\ regions
conveniently located at 
$\lambda\lambda 2026, 2062$\,\AA\ (\ZnII)
and $\lambda\lambda 2056, 2062, 2066$\,\AA\ (\CrII).
All of these factors account for the fact that 
\ZnII\ and \CrII\ absorption lines from DLAs have been 
the target of many studies since the 1990s, even though echelle
spectrographs on 8-10\,m class telescopes now afford
a more comprehensive assessment of the overall chemical
composition of QSO absorbers (e.g. Prochaska et al. \cite{prochaska01}).
Accordingly, in this paper we focus on the abundances
of Zn and Cr in CORALS DLAs and compare them with the
large body of such measurements now available for optically
selected DLAs.

The paper is organised as follows. In \S2\ we describe the 
observations, data reduction process and column density 
measurements, while in \S3\ we list the abundances measured 
in the DLAs in our sample.  We compare the CORALS Zn and Cr 
abundances to previous surveys in \S4\ and discuss our 
findings, together with our conclusions in \S5.

\section{Observations and Data Reduction}
\label{Obs}

\begin{table*}
\begin{center}
\caption[]{Details of the QSOs observed.}
\label{table1}
\setlength{\tabcolsep}{0.10cm}
\begin{tabular}{ccclccrr}
\noalign{\smallskip}
\hline
\noalign{\smallskip}
QSO & $\it{B}$ & $z_{\rm em}$ & $z_{\rm abs}$ & Telescope/ & Resolution & Integration & ~S/N$^a$ \\
    & mag.     &              &               & Instrument &   (\AA)    & Time (s)    &     \\
(1) &   (2)    &     (3)      & (4)           &    (5)     &    (6)     & (7)         & (8) \\
\noalign{\smallskip}
\hline
\noalign{\smallskip}
B0335$-$122 & 21.5$^{b}$               & 3.442 & 3.17995 & VLT 2/UVES      & 0.20& 16\,300 & 7 \\
B0347$-$211 & \multicolumn{1}{l}{20.9} & 2.944 & 1.947   & VLT 2/UVES      & 0.14&  8\,900 & 6 \\
B0405$-$331 & \multicolumn{1}{l}{19.4} & 2.570 & 2.56932 & Keck II/ESI     & 1.16& 3\,600  & 15 \\
B0432$-$440 & \multicolumn{1}{l}{19.8} & 2.649 & 2.30205 & VLT 2/UVES      & 0.16& 14\,400 & 12 \\
B0438$-$436 & \multicolumn{1}{l}{20.7} & 2.863 & 2.34736 & VLT 2/UVES      & 0.16& 14\,700 & 25 \\
B0537$-$286 & \multicolumn{1}{l}{19.8} & 3.110 & 2.9746  & Keck I/LRIS     & 3.87& 8\,600  & 39 \\
B0913$+$003 & 21.7$^{b}$               & 3.074 & 2.7434  & Keck II/ESI     & 1.21&  3\,600 & 8 \\
B0933$-$333 & \multicolumn{1}{l}{19.7} & 2.906 & 2.6823  & VLT 2/UVES      & 0.18& 14\,400 & 20 \\
B1055$-$301 & \multicolumn{1}{l}{19.4} & 2.523 & 1.90350 & VLT 2/UVES      & 0.14& 10\,800 & 15 \\
B1228$-$113 & \multicolumn{1}{l}{20.9} & 3.528 & 2.19289 & VLT 2/UVES      & 0.15& 18\,000 & 15 \\
B1230$-$101 & \multicolumn{1}{l}{19.7} & 2.394 & 1.93136 & Magellan I/MIKE & 0.25& 18\,000 & 11 \\
B1354$-$107 & \multicolumn{1}{l}{19.0} & 3.006 & 2.50086 & VLT 2/UVES      & 0.17& 14\,400 & 21 \\
            &                          &       & \&\ 2.96682 &             & 0.19&        &  14 \\
B1418$-$064 & \multicolumn{1}{l}{20.4} & 3.689 & 3.44828 & VLT 2/UVES      & 0.23& 14\,400 & 11 \\
B2311$-$373 & \multicolumn{1}{l}{19.0} & 2.476 & 2.18210 & VLT 2/UVES      & 0.15& 10\,800 & 25 \\

\hline
\end{tabular}
\end{center}
$^{a}$ Signal to noise ratio in the wavelength region encompassing the 
\ZnII\ and \CrII\ absorption lines.\\
$^{b}$ These two values are taken from Table 3 of Ellison et al. (\cite{ellison01}).
\end{table*}

The CORALS sample of QSOs found to have DLAs with absorption 
redshifts $1.8~\leq~z_{\rm abs}~\leq~z_{\rm em}$ 
consists of a total of 18 QSOs (22 DLAs; Ellison et al. \cite{ellison01}).  
Prior to this study, spectra of sufficiently 
high resolution and signal-to-noise ratio (S/N) 
for abundance determinations had already been obtained (either 
by us or by others) for five of these DLAs and have 
appeared in the literature.  
We therefore describe here only observations 
made on the remainder of the CORALS sample, details of which 
are collected in Table 1.

$B$-band magnitudes in column (2) have been taken from Table 2 
of the recent study of Ellison et al. (\cite{ellison05a}) 
except where indicated.
Emission redshifts $z_{\rm em}$
in column (3) have been reproduced directly 
from Table 3 of Ellison et al. (\cite{ellison01}),
while the values of DLA absorption redshift, $z_{\rm abs}$
listed in column (4)
were determined from the observations presented here 
and are quoted to the precision of our measurements.
One QSO from the CORALS sample, B1251$-$407, 
is too faint ($B = 23.7$) for high resolution 
spectroscopy and no abundance measurements are therefore 
available for the two DLAs identified by Ellison et al. 
in its spectrum.  However, we do not expect this small 
gap in our survey to affect the conclusions of the present 
study for two reasons.  First, the two DLAs in B1251$-$407 
are at higher redshifts than the rest
of the CORALS sample ($z_{\rm abs} = 3.533$ and 3.752). 
Second, their neutral hydrogen column densities, 
$N$(\HI)\,$= 4 \times 10^{20}$ and $2 \times 10^{20}$,
are at the lower end of the DLA column density distribution.
Thus, the metal and dust content of these two DLAs, 
even if they turned out to be significantly different 
from those of the rest of the sample,
would have only a minor impact on integrated quantities
such as the column density weighted metallicity considered
in \S4. Specifically, if we assume that, unlike most DLAs,
these two systems have a solar abundance of Zn, the column density
weighted mean metallicity of the CORALS sample would be increased by 
only 0.1\,dex.

\subsection{Data Acquisition}

The observations were made between April 2002 and March 2004 
during several runs on a range of telescopes and instruments.  
The majority of the data were obtained in service mode using 
the Ultraviolet and Visual Echelle Spectrograph (UVES; 
Dekker et al. \cite{dekker00}) on the European Southern Observatory 
Very Large Telescope (VLT).
The spectra of B0405$-$331 and B0913$+$003 were obtained with 
the Echellette Spectrograph and Imager (ESI; Sheinis et al. \cite{sheinis02}) 
at the cassegrain focus of the Keck II telescope; for 
B0537$-$286 we used the Low-Resolution Imaging Spectrometer 
(LRIS; Oke et al. \cite{oke95}) at the cassegrain focus of 
Keck I.  B1230$-$101 was observed with the newly-commissioned 
Magellan Inamori Kyocera Echelle spectrograph 
(MIKE; Bernstein et al. \cite{bernstein03}) on the 6.5m 
Magellan Baade telescope.

Instrument settings were chosen for each QSO observation so as 
to provide coverage at the wavelengths of the redshifted \ZnII\ 
and \CrII\ absorption lines.  We achieved spectral resolutions 
varying between 0.14 and 0.25\,\AA\ FWHM for the UVES and MIKE 
data, and of $\sim$1.16\,\AA\ and 3.87\,\AA\ FWHM for the ESI 
and LRIS data respectively. 
These correspond to velocity resolutions of $\sim$7\,\kmprs, 
$\sim$13\,\kmprs, $\sim$48\,\kmprs\ and $\sim$140\,\kmprs\ for 
the UVES, MIKE, ESI and LRIS spectra respectively.

\subsection{Data Reduction}

Due to the differing nature of the instruments used, the spectra 
were not identically reduced, although the same standard steps 
were incorporated in each case. The reduction of the UVES data 
was performed with the ESO UVES data reduction pipeline, while 
the other spectra were reduced using standard IRAF\footnote{IRAF 
is distributed by the National Optical Astronomy Observatories, 
which are operated by the Association of Universities for Research 
in Astronomy, Inc., under cooperative agreement with the National 
Science Foundation} routines.  
The two-dimensional images were trimmed and the bias level 
subtracted using the over-scan regions.  Pixel-to-pixel variations 
were corrected for by dividing through by a normalized flat-field 
of high S/N that had been produced by co-adding several flat-field 
exposures.  Pixels which had been affected by cosmic ray hits were 
corrected for, where possible, and the echelle orders or single 
long-slit spectrum were traced.  The one-dimensional spectrum was 
then extracted and the sky spectrum subtracted (in the case of MIKE 
taking into account the slit tilt, which is not perpendicular to 
each order and varies over the CCD).  

The individual sky-subtracted spectra were then wavelength 
calibrated (using comparison lamp spectra typically taken 
immediately following and/or prior to each science frame and 
extracted in the same way as above, except for sky subtraction) and 
then corrected to a vacuum heliocentric wavelength scale.  In the 
next step, the spectra of each object were rebinned to a common 
linear wavelength scale, with a bin size close to their original size 
and co-added, rejecting any remaining cosmic rays or bad pixels.
The co-addition was performed without any weighting and the 
corresponding error spectra were summed in quadrature.  Gaussian 
fits to several emission lines from the reference lamp spectra near 
to the redshifted \ZnII\ and \CrII\ lines were made to estimate the 
values of the spectral resolution (FWHM) listed in Table 1.

\subsection{Spectral Line Fitting}

\begin{figure*}
\vspace*{-1.1cm}
\centerline{\resizebox{20cm}{!}{\includegraphics[angle=270]{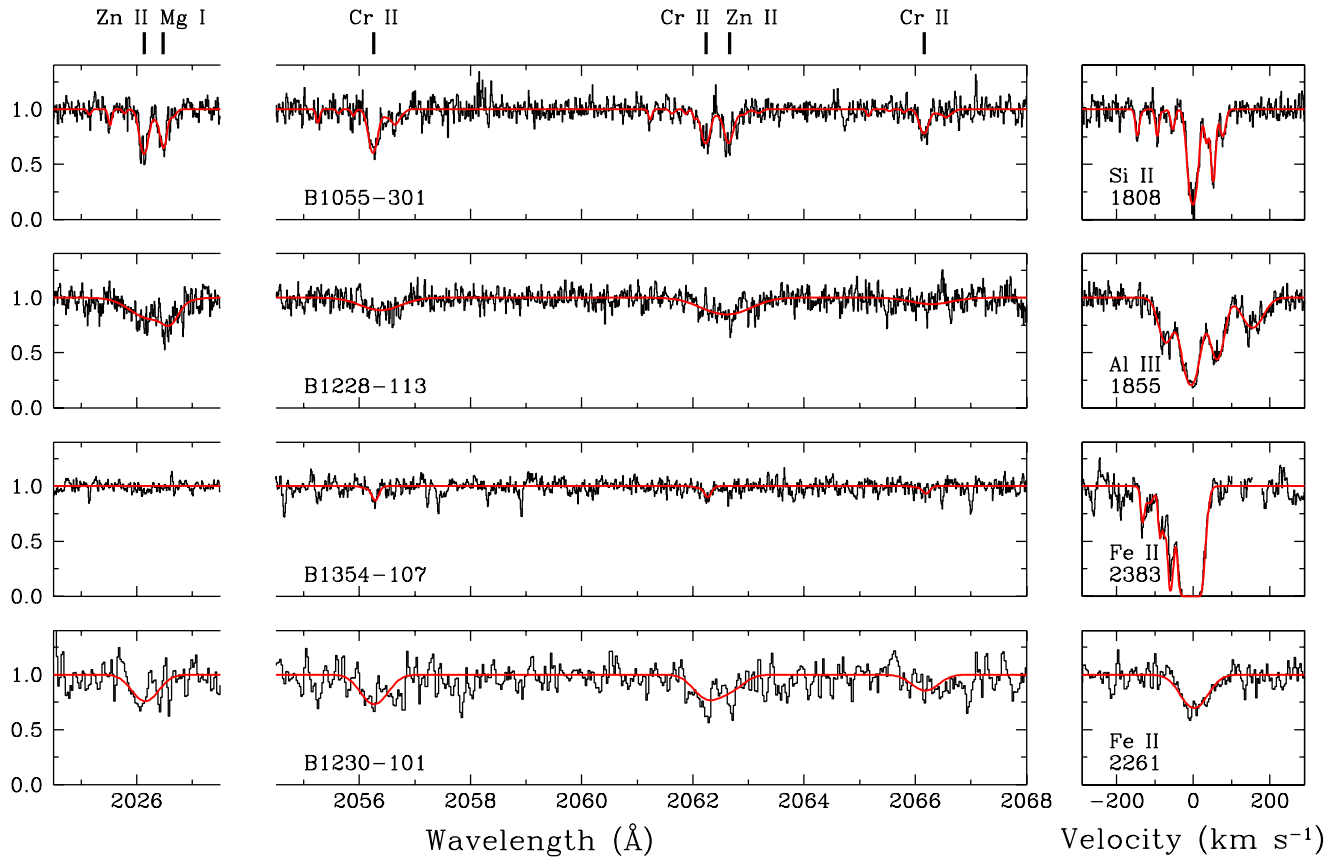}}}
\vspace{-2.4cm}
\caption{
Sample portions of spectra of CORALS QSOs, showing the
\ZnII~$\lambda 2026.14$ and \MgI~$\lambda 2026.48$ lines ({\it left}), 
and the \CrII~$\lambda \lambda 2056.26$, 2062.24, 2066.16 
and \ZnII~$\lambda 2062.66$ lines
({\it centre}) in four DLAs. 
In the $z_{\rm abs} = 2.50086$ DLA in the spectrum of B1354$-$107 
(third panel), we just detect \CrII~$\lambda \lambda2056.26$, 2062.24,
while the other lines are below our detection limit. 
In the far right panels we have reproduced examples of stronger
transitions from other elements which were used to 
determine the velocity structure of the gas in each DLA.
The continuous coloured lines show the profile fits to the data
produced by VPFIT (see text).
}
\label{fig1}
\end{figure*}

Metal absorption lines associated with the DLA systems in each 
QSO spectrum were initially identified based on the redshifts 
reported by Ellison et al. (\cite{ellison01}).  The vacuum-heliocentric 
wavelengths measured from our spectra for the strongest component 
in each absorption system were then used to determine a more accurate 
redshift for the DLA system, based on the average of the redshifts 
calculated from each (non-saturated) line.  The spectra were then 
reduced to a rest-frame wavelength scale and cut into sections 
corresponding to $\pm~2000\,\kmprs$ around each absorption line.  
These sections were then normalised to the local continuum by 
division by a spline fit to portions of the spectrum judged to 
be free of absorption.  

\begin{table}
\begin{center}
\caption[]{Transition wavelengths and $\it{f}$-values used in this study.}
\label{table2}
\setlength{\tabcolsep}{0.10cm}
\begin{tabular}{ccl}
\noalign{\smallskip}
\hline
\noalign{\smallskip}
Species & Wavelength & \multicolumn{1}{c}{$\it{f}$} \\
        &  (\AA)     &          \\
\noalign{\smallskip}
\hline
\noalign{\smallskip}
\ZnII\  & 2026.1370  & 0.501   \\
\MgI\   & 2026.4768  & 0.113   \\
\CrII\  & 2056.2569  & 0.103   \\
\CrII\  & 2062.2361  & 0.0759  \\
\ZnII\  & 2062.6604  & 0.246   \\
\CrII\  & 2066.1640  & 0.0512  \\

\hline
\end{tabular}
\end{center}
\end{table}

\begin{table*}
\begin{center}

\caption[ ]{Zn and Cr abundances in the damped \lya\ systems observed.}
\label{table3}
\setlength{\tabcolsep}{0.20cm}
\begin{tabular}{cccccccc}
\noalign{\smallskip}
\hline
\noalign{\smallskip}
QSO          & $z_{\rm{DLA}}$ & $\it{N}$(H$^{0}$) & $\it{N}$(Zn$^{+}$) & $\it{N}$(Zn$^{+}$)/$\it{N}$(H$^{0}$) &  $\it{N}$(Cr$^{+}$) & $\it{N}$(Cr$^{+}$)/$\it{N}$(H$^{0}$) &  $\it{N}$(Cr$^{+}$)/$\it{N}$(Zn$^{+}$)  \\
             &             & (10$^{20}$cm$^{-2}$) & (10$^{12}$cm$^{-2}$) & (10$^{-9}$) & (10$^{12}$cm$^{-2}$) & (10$^{-9}$) &   \\
(1) & (2) & (3) & (4) & (5) & (6) & (7) & (8)  \\                  
\hline
\noalign{\smallskip}
B0335$-$122  &3.17995   &  6.0 $\pm$ 1.5  & $\leq$ 1.8    & $\leq$ 3.0    &  $^{a}$       & $^{a}$       &  \ldots \\
\noalign{\smallskip}
B0347$-$211  &1.947     &  2.0 $\pm$ 0.5  & $\leq$ 2.4    & $\leq$ 12     &  $\leq$ 7.4   & $\leq$ 37    &  \ldots \\
\noalign{\smallskip}
B0405$-$331  &2.56932   &  4 $\pm$ 1      & $\leq$ 5.5    & $\leq$ 14     &  $\leq$ 21    & $\leq$ 54    &  \ldots \\ 
\noalign{\smallskip}
B0432$-$440  &2.30205   &  6.0 $\pm$ 1.5  & $\leq$ 1.6    & $\leq$ 2.7    &  13 $\pm$ 2   & 22 $\pm$ 6   &  $\geq$ 8.4 \\
\noalign{\smallskip}
B0438$-$436  &2.34736   &  6.0 $\pm$ 1.5  & 5.3 $\pm$ 0.3 & 8.8 $\pm$ 2.3 &  12 $\pm$ 1.4 & 19 $\pm$ 5   &  2.2 $\pm$ 0.3\\
\noalign{\smallskip}
B0537$-$286  &2.9746    &  2.0 $\pm$ 0.5  & $\leq$ 3.4    & $\leq$ 17     &  $\leq$ 20    & $\leq$ 100   &  \ldots \\
\noalign{\smallskip}
B0913$+$003  &2.7434    &  5.5 $\pm$ 1.4  & $\leq$ 6.6    & $\leq$ 12     &  $\leq$ 29    & $\leq$ 53    &  \ldots \\
\noalign{\smallskip}
B0933$-$333  &2.6823    &  3.0 $\pm$ 0.8  & $\leq$ 0.98   & $\leq$ 3.3    &  $\leq$ 8.7   & $\leq$ 29    &  \ldots \\
\noalign{\smallskip}
B1055$-$301  &1.90350   &  35  $\pm$ 9    & 8.2 $\pm$ 0.5 & 2.3 $\pm$ 0.6 &  41 $\pm$ 2   & 12 $\pm$ 3   &  5.0 $\pm$ 0.4\\
\noalign{\smallskip}
B1228$-$113  &2.19289   &  4   $\pm$ 1    & 10.3 $\pm$ 0.8& 26 $\pm$ 7    &  28 $\pm$ 2   & 69 $\pm$ 18  &  2.7 $\pm$ 0.3\\
\noalign{\smallskip}
B1230$-$101  &1.93136   &  3.0 $\pm$ 0.8  & 8.7 $\pm$ 1.0 & 29 $\pm$ 8    &  47 $\pm$ 4   & 160 $\pm$ 40 &  5.4 $\pm$ 0.8\\
\noalign{\smallskip}
B1354$-$107  &2.50086   &  2.5 $\pm$ 0.6  & $\leq$ 0.5    & $\leq$ 2      & 7.4 $\pm$ 0.8 & 29 $\pm$ 8   &  $\geq$ 15\\
\noalign{\smallskip}
B1354$-$107  &2.96682   &  6.0 $\pm$ 1.5  & $\leq$ 0.85   & $\leq$ 1.4    &  $\leq$ 6.0   & $\leq$ 10    &  \ldots \\
\noalign{\smallskip}
B1418$-$064  &3.44828   &  2.5 $\pm$ 0.6  & $\leq$ 0.96   & $\leq$ 3.9    &  $\leq$ 5.1   & $\leq$ 20    &  \ldots \\
\noalign{\smallskip}
B2311$-$373  &2.18210   &  3.0 $\pm$ 0.8  & $\leq$ 0.66   & $\leq$ 2.2    &  $\leq$ 3.4   & $\leq$ 11    &  \ldots \\

\hline
\end{tabular}

\end{center}
$^{a}$ The \ion{Cr}{ii} lines were not covered by the 
UVES observations of this QSO as they fell in the 
wavelength gap between the two red-arm CCDs .

\end{table*}

Ion column densities were deduced from the observed absorption 
lines using the line fitting software package 
VPFIT\footnote{VPFIT is available from 
http://www.ast.cam.ac.uk/$\sim$rfc/ vpfit.html}. VPFIT fits 
multiple Voigt profiles (convolved with the instrument 
profile) to absorption line components.  Initial guesses 
for the column density, redshift and Doppler $\it{b}$ 
parameter for each absorption component are used as inputs 
into the program.  VPFIT then computes the $\chi^{2}$ 
goodness-of-fit (taking into account the error on each pixel) 
and automatically varies the parameters, iterating until 
$\chi^{2}$ has been minimised and the best fit has been found.  
Components were rejected from the fit if their Doppler 
parameters were lower than approximately half the spectral 
resolution. For the rest-frame wavelengths and $f$-values of
the transitions, which are inputs to VPFIT, we consulted the 
recent compilation by Morton (\cite{morton03}); for reference, 
values for the \ZnII\ and \CrII\ multiplets are reproduced 
in Table 2.

The first step in the line fitting procedure
was to determine the multi-component velocity structure
of gas in the damped system by fitting lines spanning
a wide range of $f$-values, preferably multiplets 
of \FeII\ and \SiII.
The parameters of the fit---that is redshift and Doppler
parameter---were then kept fixed and 
applied to the generally weaker \ZnII\ and \CrII\ lines.
In cases where only some of the \FeII\ and \SiII\
absorption components are detected in \ZnII\ and \CrII, 
the total column densities of Zn$^+$ and Cr$^+$ were
adjusted upwards to allow for the unseen components
in the same proportion as determined for \FeII\
and \SiII.  
The implicit assumption is that the relative proportions
of absorbers in each component are the same for all the ions
considered. While in principle this may not be true, because
of possible differences in the degree of dust depletion of
different elements, in practice element ratios are found to
be remarkably uniform between the multiple velocity components
within a DLA (e.g. Prochaska 2003).
The correction for unseen components amounted to less than $\sim 10$\%  
of the total column density in all but two cases, 
B0432$-$440 and B0933$-$333, where the correction
for unseen components is as much as $\sim 50$\%.
Indeed, this correction for undetected components
was the main reason why we used VPFIT to analyse the 
\ZnII\ and \CrII\ absorption. In all cases considered
here, the \ZnII\ and \CrII\ lines 
are unsaturated and we would
have obtained the same values of 
$N$(Zn$^+$) and $N$(Cr$^+$)---apart from
the `incompleteness' correction---had we analysed them
with the optical depth method (Hobbs \cite{hobbs74})
which some prefer to profile fitting. 

\begin{table}
\begin{center}
\caption[]{Velocity dispersion parameters of the strongest 
component of detected Zn II and Cr II lines}
\label{table4}
\setlength{\tabcolsep}{0.20cm}
\begin{tabular}{ccc}
\noalign{\smallskip}
\hline
\noalign{\smallskip}
QSO & $z_{\rm{DLA}}$ &  $b$ \\
    &                & (\kmprs) \\
\noalign{\smallskip}
\hline
\noalign{\smallskip}
B0432$-$440 & 2.30205 & 12 \\
B0438$-$436 & 2.34736 &  8 \\
B1055$-$301 & 1.90350 & 16 \\
B1228$-$113 & 2.19289 & 73 \\
B1230$-$101 & 1.93136 & 47 \\
B1354$-$107 & 2.50086 & 12 \\
\hline
\end{tabular}
\end{center}
\end{table}

\begin{table*}
\begin{center}
\caption[ ]{Abundances in the CORALS DLA survey.  Measurements are given in 
logarithmic units relative to the solar abundances from 
Lodders (\cite{lodders03}): 
12 + log(Zn/H) = 4.63, 12 + log(Cr/H) = 5.65, 12 + log(Fe/H) = 7.47, 12 + log(Si/H) = 7.54.
}
\label{table5}
\setlength{\tabcolsep}{0.20cm}
\begin{tabular}{cccccccc}
\noalign{\smallskip}
\hline
\noalign{\smallskip}
QSO & $z_{\rm{abs}}$ & $\it{N\rm{(H^{0})}}$& [Zn/H] & [Cr/H] & [Fe/H] & [Si/H] & Ref. \\
    &                & (10$^{20}$cm$^{-2}$)&        &        &        &        &     \\
\noalign{\smallskip}
\hline
\noalign{\smallskip}
B0335$-$122 & 3.17995 & 6.0 & $\leq$ $-$1.16 & \ldots         & $-$2.61 & $-$2.56&    \\
B0347$-$211 & 1.947   & 2.0 & $\leq$ $-$0.54 & $\leq$ $-$1.08 & $\leq$ $-$1.16 & \ldots&      \\
B0405$-$331 & 2.56932 & 4.0 & $\leq$ $-$0.49 & $\leq$ $-$0.92 & $-$1.74 & $-$1.40 &     \\
B0432$-$440 & 2.30205 & 6.0 & $\leq$ $-$1.21 & $-$1.30        & $-$1.45 & $-$1.12 &     \\
B0438$-$436 & 2.34736 & 6.0 & $-$0.68        & $-$1.36        & $-$1.30 & \ldots &     \\
B0458$-$020 & 2.0395  & 44.7 & $-$1.15       & $-$1.50        & $-$1.61 & $\geq$ $-$2.10 &  1, 2   \\
B0528$-$250 & 2.141   & 8.9 & $-$1.45        & $-$1.50        & $-$1.57 & $-$1.27 &  3, 4   \\
B0528$-$250 & 2.811   & 12.9 & $-$0.47       & $-$1.11        & $-$1.11 & $-$0.64 &  3, 4   \\
B0537$-$286 & 2.9746  & 2.0 & $\leq$ $-$0.40 & $\leq$ $-$0.65 & \ldots  & \ldots &     \\
B0913$+$003 & 2.7434  & 5.5 & $\leq$ $-$0.55 & $\leq$ $-$0.92 & $-$1.59 & $-$1.47 &    \\
B0933$-$333 & 2.6823  & 3.0 & $\leq$ $-$1.12 & $\leq$ $-$1.19 & $-$1.54 & $-$1.22 &     \\
B1055$-$301 & 1.90350 & 35.0 & $-$1.26       & $-$1.58        & $-$1.57 & $-$1.13 &     \\
B1228$-$113 & 2.19289 & 4.0 & $-$0.22        & $-$0.81        & \ldots  & \ldots &     \\
B1230$-$101 & 1.93136 & 3.0 & $-$0.17        & $-$0.45        & $-$0.63 & $-$0.24 &     \\
B1354$-$107 & 2.50086 & 2.5 & $\leq$ $-$1.32 & $-$1.18        & $-$1.25 & \ldots &     \\
B1354$-$107 & 2.96682 & 6.0 & $\leq$ $-$1.48 & $\leq$ $-$1.65 & $-$1.54 & $-$1.31 &     \\
B1418$-$064 & 3.44828 & 2.5 & $\leq$ $-$1.04 & $\leq$ $-$1.34 & $-$1.72 & $-$1.48 &     \\
B2311$-$373 & 2.18210 & 3.0 & $\leq$ $-$1.29 & $\leq$ $-$1.60 & $-$1.70 & $-$1.52 &     \\
B2314$-$409 & 1.8573  & 7.9 & $-$1.01        & $-$1.17        & $-$1.29 & $-$1.03 &  5   \\
B2314$-$409 & 1.8745  & 1.3 & $\leq$ $-$1.18 & $\leq$ $-$1.59 & $-$1.85 & $-$1.86 &  5   \\
\hline
\end{tabular}
\end{center}
References:
1 - Prochaska \&\ Wolfe (\cite{prochaska99}),  
2 - Prochaska et al. (\cite{prochaska01}), 
3 - Centurion et al. (\cite{centurion03}), 
4 - Lu et al. (\cite{lu96}), 
5 - Ellison \&\ Lopez (\cite{ellison01a}).
\end{table*}

Out of the 15 DLAs listed in Table 1, we report
positive detections of both the \ZnII\ and \CrII\ multiplets
in four cases (see Table 3).  
In two additional cases we detect \CrII\ but not \ZnII,
while in the remaining nine DLAs we place upper limits
to both $N$(Zn$^+$) and $N$(Cr$^+$).
The upper limits are $3\,\sigma$, deduced from the measured
S/N ratios (listed in column (8) of Table 1) and the 
velocity spread of the absorption implied by stronger
lines of \FeII\ and \SiII, as explained above.

Figure 1 shows examples of the spectra in the regions
of the \ZnII\ and \CrII\ multiplets, including three 
of the \ZnII\ and \CrII\ detections and one case when only
\CrII\ is detected.\footnote{Spectra of all the CORALS QSOs
covered in the present study are available from 
ftp://ftp.ast.cam.ac.uk/pub/papers/CORALS/ }  
Also shown in the figure are examples 
of the stronger lines used to determine the velocity structure
of each DLA, as well as the fits generated by VPFIT.
Table 4 lists values of the Doppler $b$ parameter for the
detected Zn II and Cr II lines; full details of the velocity
structure of other absorption lines will be presented in
a future publication (Akerman 2005).

\section{Abundance determinations}

Table 3 lists the column densities of \ZnII\ and \CrII,
or upper limits, measured in this study together 
with the 1\,$\sigma$ uncertainties in their values. 
Since the first ions are the dominant ionization 
stages of Zn and Cr in \HI\ regions, we can readily
derive the abundances of these two elements by dividing
the values of $N$(Zn$^+$) and $N$(Cr$^+$) by the 
neutral hydrogen column densities $N$(H$^0$) listed
in column (3) of Table 3---the results are given 
in columns (5) and (7) respectively.
The values of $N$(H$^0$) are reproduced directly
from the compilation by Ellison et al. (\cite{ellison01}) because
(a) few of our spectra include the \lya\
absorption line and (b) even in cases where we 
cover the transition, the intermediate resolution
spectra of Ellison et al. are better suited than ours
to the determination of $N$(H$^0$) by fitting the damping
wings of the \lya\ absorption line. 
We have adopted throughout a conservative
error of $\pm 25$\% to $N$(H$^0$), based 
on the typical accuracy with which 
such measurements have been reported in the
literature.

In Table 5 we list the abundances of Zn, Cr, Fe, and Si, 
expressed in logarithmic units relative to 
solar, for each of the DLAs in the complete CORALS sample;
the list includes all the new measurements reported here
plus the five already available from 
the literature---references are given
in the last column of Table 5. 
The only DLAs in the compilation by Ellison et al. (\cite{ellison01}) 
which are missing from our sample are 
the two systems in front of the very faint QSO B1251$-$407,
as explained in \S2.
We have adopted throughout the solar abundances 
proposed by Lodders (\cite{lodders03}) in her comprehensive 
reassessment of meteoritic and photospheric abundances; those from
the more recent compilation
by Asplund, Grevesse, \& Sauval (\cite{asplund04}) 
differ by no more than 0.03\,dex for the elements
of interest here. All the Zn and Cr 
measurements collected in Table 5 (and indeed those
of the larger comparison sample discussed in \S4)
have been reduced to the same set of $f$-values,
as listed in Table 2, except for cases where the
differences in column density would have 
amounted to less than a few percent.\footnote{Most
recent work has used the same set of $f$-values for
the \ZnII\ and \CrII\ multiplets, from the laboratory
measurements by Bergeson \& Lawler (\cite{bergeson93}). Small
differences between the values quoted by different
observers result from rounding errors.} It was not possible,
however, to ensure the same degree of homogeneity
for the \FeII\ and \SiII\ measurements.


\begin{figure*}
\vspace{-3cm}
\centerline{\resizebox{18cm}{!}{\includegraphics[angle=270]{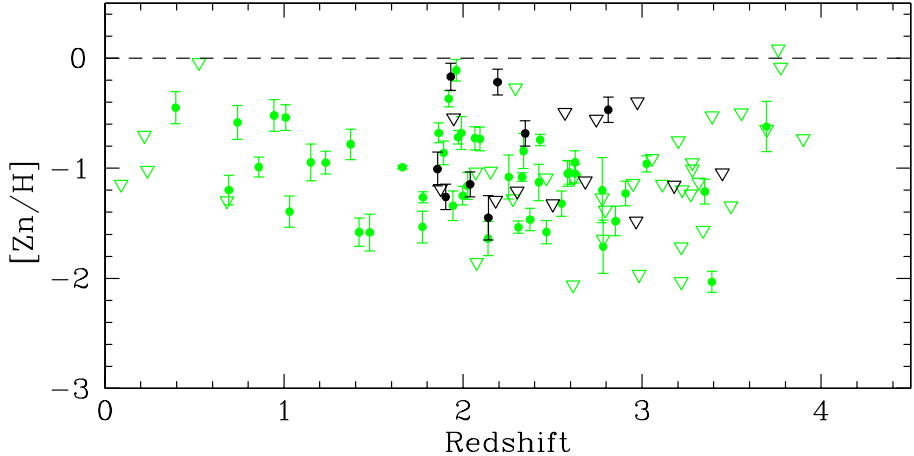}}}
\vspace{-3.5cm}
\caption{
Comparison of the abundance of Zn 
measured in DLAs in the CORALS sample (black)
and in the control sample by Kulkarni et al. \cite{kulkarni05} (coloured).
Upper limits, corresponding to non-detections
of the \ZnII\ doublet, are indicated by open
triangles. The abundance of Zn is plotted on a logarithmic
scale relative to the solar value shown by the broken line
at [Zn/H]\,=\,0.0.
}
\label{fig2}
\end{figure*}

\section{Metallicity and dust in the CORALS survey}

The purpose of this study is to test whether element 
abundances in CORALS DLAs are significantly different 
from those of DLAs drawn from optically selected QSO samples, 
and therefore to determine the extent, if any, to which 
previous surveys have been biased against metal-rich, 
high column density absorbers.  We consider this question
from two points of view, by examining first the metallicity
distributions indicated by the [Zn/H] determinations,
and then the degree of depletion of refractory elements
implied by the [Cr/Zn] measures.

\begin{figure}
\vspace{-2cm}
\centerline{\resizebox{12cm}{!}{\includegraphics[angle=270]{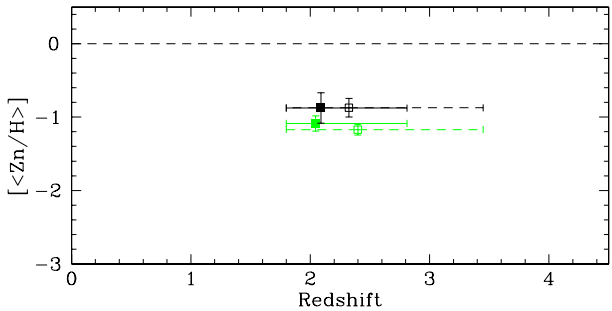}}}
\vspace{-2.25cm}
\caption{
Column density-weighted metallicities of DLAs in the 
CORALS (black) and Kulkarni et al. \cite{kulkarni05} (coloured) samples.  
For each sample we show two values of 
${\rm [} \langle {\rm Zn/H} \rangle {\rm ]}$;
the solid squares correspond to the values obtained by
considering only \ZnII\ detections, while the open
squares (and dashed error bars) are obtained if the 
upper limits to [Zn/H] in Figure 2
are included as if they were detections.
The squares are plotted at the median redshift 
of the DLAs in each sample.
}
\label{fig3}
\end{figure}

\subsection{Comparison of Zn Abundances}

The most extensive, homogeneous, compilation of [Zn/H]
measurements in DLAs has recently been assembled by Kulkarni
et al. (2005); it includes 51 detections and 36 upper
limits over the redshift interval $z_{\rm abs} = 0.09$ to 3.90.
(Three of the \ZnII\ detections in the Kulkarni et al. data
set are in common with the CORALS survey and have thus not been 
included in the control sample in the following analysis.)
Although larger compilations of DLA abundance measurements
have been published (e.g. Prochaska et al. 2003), 
they bring together data for other elements, 
mostly Si and Fe, in addition to Zn. 
The improved statistics
afforded by such compilations are offset, in our view,
by systematic uncertainties due to differing degrees
of dust depletion and possible nucleosynthetic departures
from solar relative abundances of different elements.
Thus, we consider it more appropriate to restrict the
present analysis to the comparison of Zn abundances
alone.

The CORALS [Zn/H] abundances are compared with those
from the compilation by Kulkarni et al. (\cite{kulkarni05}) in 
Figure 2 (after rescaling the latter to the solar
value from Lodders \cite{lodders03}). 
From a visual inspection of the plot we conclude that:
(a) both the CORALS and the comparison sample of
optically selected DLAs are generally metal-poor,
with typical values of [Zn/H] well below solar; and
(b) there is a hint that, overall, the CORALS DLAs
may have marginally higher metallicities, although
the considerable number of upper limits 
complicates the comparison.
We now address these points quantitatively.

\begin{figure*}
\vspace{-3cm}
\centerline{\resizebox{18cm}{!}{\includegraphics[angle=270]{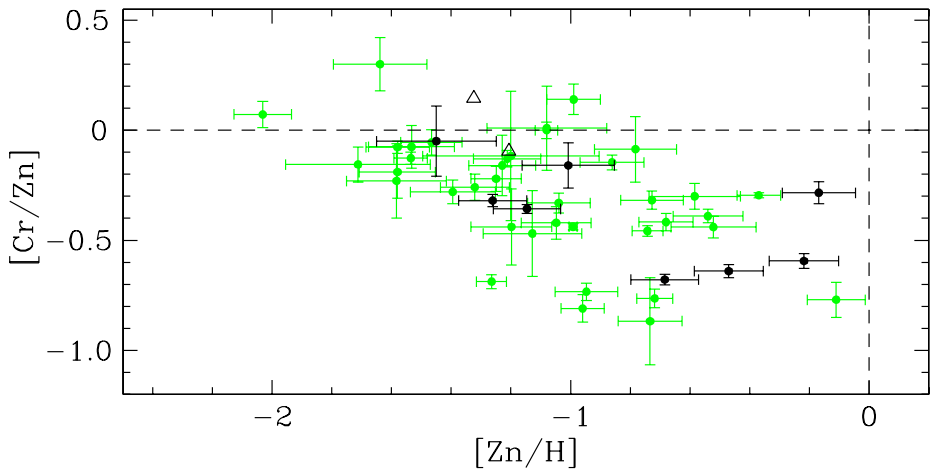}}}
\vspace{-3.5cm}
\caption{Cr/Zn ratio against Zn abundance for CORALS DLAs (black) 
and DLAs from the compilation by Kulkarni et al. (\cite{kulkarni05}; coloured). 
Lower limits, corresponding to cases where \ion{Cr}{ii}
lines are detected by \ion{Zn}{ii} lines are not, are indicated by open
triangles. Abundance ratios are plotted in logarithmic units 
relative to the solar values shown by the broken lines at 
[Cr/Zn] = 0.0 and [Zn/H] = 0.0.
}
\label{fig4}
\end{figure*}

The quantity which is of interest for `cosmic'
chemical evolution models (Pei \& Fall \cite{pei95})
is the column density-weighted metallicity

\begin{equation}
    {\rm [} \langle{\rm (Zn/H)}_{\rm DLA}\rangle {\rm ]} = 
    \log \langle{\rm (Zn/H)}_{\rm DLA}\rangle - \log {\rm (Zn/H)}_{\odot}
    \label{}
\end{equation}
\noindent where
\begin{equation}
        \langle{\rm (Zn/H)}_{\rm DLA}\rangle = 
        \frac{\sum\limits_{i=1}^{n} N{\rm(Zn}^+{\rm)}_i}
        {\sum\limits_{i=1}^{n} N{\rm(H}^0{\rm)}_i} 
        \label{}
\end{equation}

\noindent which is a measure of the degree of metal enrichment 
of the DLA population as a whole. The summation in eq. (2) is
over the $n$ DLAs considered in a given sample. 
The present sample is too small to consider subsets of the 
data within different redshift intervals, as done by 
Pettini et al. (\cite{pettini99}), Prochaska et al. 
(\cite{prochaska03}), and Kulkarni et al. (\cite{kulkarni05}).
Instead we compute eq. (2) for the CORALS sample as a whole,
and compare the result with the analogous quantity for the
Kulkarni et al. (\cite{kulkarni05}) sample, {\em computed 
over the same redshift interval spanned by the CORALS data}, 
$z_{\rm abs} = 1.86 - 3.45$  for the full set (57 DLAs), 
and $z_{\rm abs} = 1.86 - 2.81$ 
for the \ZnII\ detections only (27 DLAs).
We find 
${\rm [} \langle{\rm (Zn/H)}_{\rm DLA}\rangle {\rm ]} = -0.88 \pm 0.21$
for CORALS DLAs, and
${\rm [} \langle{\rm (Zn/H)}_{\rm DLA}\rangle {\rm ]} = -1.09 \pm 0.10$
for the control sample of Kulkarni et al. (\cite{kulkarni05}).
These values, which differ at only the $1\,\sigma$ level,
were calculated considering only the \ZnII\ detections. 
Repeating the calculations, but now including the upper
limits as if they were detections, we obtain 
${\rm [} \langle{\rm (Zn/H)}_{\rm DLA}\rangle {\rm ]} = -0.87 \pm 0.13$
and $-1.17 \pm 0.07$ respectively.
The errors were estimated using bootstrap techniques 
(Efron \&\ Tibshirani \cite{efron93}). For each sample
(CORALS and the control sample), we constructed random
datasets by drawing $n$ times from the real dataset 
with replacement, where $n$ is the number of 
[Zn/H] measurements in each sample.
The procedure was repeated a million times to build
up a distribution of values of 
${\rm [} \langle{\rm (Zn/H)}_{\rm DLA}\rangle {\rm ]}$;
we quote the standard deviation of this distribution
as our estimate of the error in the quantity
${\rm [} \langle{\rm (Zn/H)}_{\rm DLA}\rangle {\rm ]}$.

These results are shown graphically in Figure 3. 
Again we see that the two samples are very similar.
The marginally higher metallicity of the CORALS sample
may be real.
Alternatively it may be an artifact of small number 
statistics or, in the case where we include upper 
limits, it may be due to the higher proportion of upper
limits skewing the results---over one half of the CORALS
measures of [Zn/H] are upper limits, compared with
one third for the control sample of Kulkarni et al. (\cite{kulkarni05}).

In order to clarify this point, we conducted a statistical 
test based on ``survival statistics'' (which takes account 
of upper limits) with the program ASURV 
(LaValley, Isobe \&\ Feigelson \cite{lavalley92}), 
which implements the statistical methods of 
Feigelson \& Nelson (\cite{feigelson85}).  
Using a Peto-Prentice test (Latta \cite{latta81}), we tested 
the null hypothesis that the two samples are drawn from the 
same parent population and found the hypothesis to be true 
at the 90\% confidence level.  We conclude that CORALS DLAs 
do not exhibit significantly different metallicities from 
those of existing, larger, samples of DLAs assembled from 
optically selected QSO surveys.

\begin{figure*}
\vspace{-2.25cm}
\centerline{\resizebox{15cm}{!}{\includegraphics[angle=270]{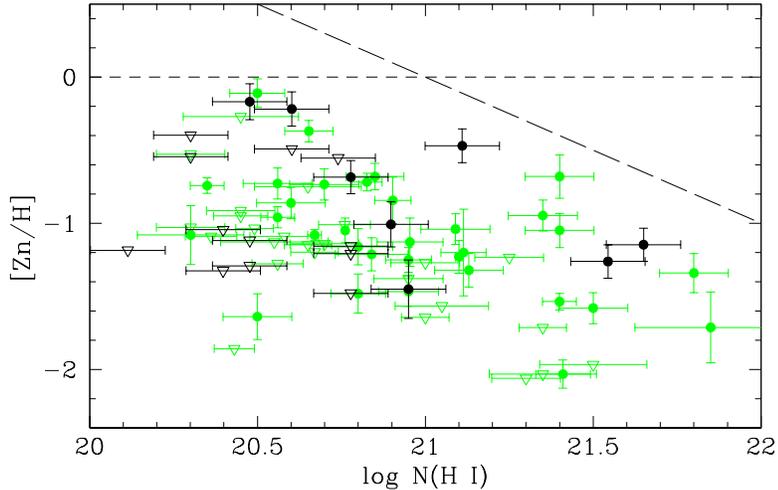}}}
\vspace{-2.5cm}
\caption{Zn abundance vs. \HI\ column density for the 20 CORALS
DLAs (black) and the 57 DLAs from the compilation by Kulkarni
et al. (2005; coloured) in the redshift range $1.86 \leq z \leq 3.45$.
Symbols have the same meaning as in Figure 2. The long-dash line  
indicates the limit ${\rm [Zn/H]} + \log N{\rm (\HI)} > 21$
beyond which Prantzos \& Boissier 
(\cite{prantzos00}) noted that there are no DLAs.
This limit seems to apply to CORALS DLAs too, although it could 
just be a reflection of the small size of
our current sample (DLAs with $\log N$(\HI)\,$ > 21$ are rare).
Alternatively, there may be reasons other than dust-induced bias
to explain this apparent `zone of avoidance'. Possibly, the
cross-section for high $N$(\HI) absorption decreases with
increasing metallicity as gas is consumed by star formation.
}
\label{fig5}
\end{figure*}

\subsection{Comparison of Cr/Zn ratios}

An indication of the degree of depletion of refractory 
elements onto dust grains may be 
obtained from the ratio of the abundances of chromium 
to zinc, as explained in \S1.  
In Figure 4 we plot this ratio against the metallicity [Zn/H]
for each of the CORALS DLAs (coloured) from Table 5 
together with analogous measurements (black) 
from the compilations by Khare et al. (\cite{khare04}) 
and Kulkarni et al. ({\cite{kulkarni05}), 
after rescaling their values to 
the same solar abundances used here.

Figure 4 shows the trend of increasing Cr depletion 
with increasing metallicity which was previously noted by 
Pettini et al. (\cite{pettini97a}) and 
shown by Prochaska \& Wolfe (2002)
to be a general feature of refractory elements in DLAs.
In systems with  [Zn/H]\,$\simlt -1.5$,
[Cr/Zn] is approximately solar---indicating 
that there is little dust depletion at such low
metallicities---while when [Zn/H]\,$ > 1$,
up to $\sim 90$\% of the Cr can be `missing' from the
gas phase and presumably be in solid form.
Even so, in none of the DLAs do we see the extreme
depletions of Cr, by two orders of magnitude, 
commonly measured in cold clouds of the 
Milky Way disk (Savage \& Sembach 1996).
At the typical DLA metallicity, 
${\rm [} {\rm (Zn/H)}_{\rm DLA} {\rm ]} \approx -1$,
approximately $1/2$ to $2/3$ of the Cr is in the dust
([Cr/Zn]\,$\approx -0.3$ to $-0.5$) 
although there is considerable dispersion in the
depleted fraction $f_{\rm Cr}$.

All of these facets of the depletion of refractory
elements in DLAs have recently been discussed by
Vladilo (2004) who linked them to the 
(generally early) chemical evolution of the galaxies 
where the absorption systems originate. 
The dependence of $f_{\rm Cr}$ on [Zn/H] may 
reflect a metallicity dependence of the
efficiency of dust formation in the ejecta of 
core-collapse supernovae and in the winds
of late-type giants. On the other hand,
the large scatter which accompanies the overall trend 
is presumably an indication of how the detailed balance
between the processes of dust formation, accretion,
and destruction is affected by the local physical 
conditions in the ISM.

As far as the present work is concerned, it is evident from
Figure 4 that the depletions of Cr in the CORALS sample conform
to the overall pattern described above. The CORALS DLAs
are not extreme in their values of $f_{\rm Cr}$, nor
do they exhibit lower values of [Cr/Zn] at a given [Zn/H]
than DLAs drawn from optically selected QSO samples.\footnote{Although 
three out of the four CORALS DLAs with 
${\rm [Zn/H]} < -1.0$ have ${\rm [Cr/Zn]} < -0.5$ compared to 
only five out of 16 from the control sample, these two fractions 
are different at less than the $1\,\sigma$ significance level even 
when using Poisson statistics (which overestimate the significance 
of any differences between samples when the number of measurements 
is so small).}
If there are any DLAs where the depletions of refractory elements 
approach the high values typical of cold clouds in the Milky Way disk, 
we have not found them yet. 
Based on the data in Figure 4, there is no evidence that
CORALS DLAs should redden the spectra of background QSOs
any more than a typical DLA from an optically selected sample.

\section{Discussion}

The work described in this paper concludes a project begun six
years ago to test the extent to which existing DLA samples are
biased by dust reddening against gas-rich galaxies of high
metallicity. The first results, reported by Ellison et al. (2001),
showed that $\Omega_{\rm DLA}$ had not been significantly
underestimated. To that conclusion we now add the findings that:

(1) At redshifts $1.9 < z < 3.5$, the metallicity of CORALS DLAs, 
as measured by the [Zn/H] ratio, 
is only marginally higher (at a statistically
{\em insignificant} level of only $1\,\sigma$) than that of 
DLAs drawn from optically selected QSOs---we determine
${\rm [} \langle{\rm (Zn/H)}_{\rm DLA}\rangle {\rm ]} = -0.88 \pm 0.21$
for CORALS DLAs, and
${\rm [} \langle{\rm (Zn/H)}_{\rm DLA}\rangle {\rm ]} = -1.09 \pm 0.10$
for the control sample of Kulkarni et al. (\cite{kulkarni05})
over the same redshift interval.  We note that none of the 
CORALS DLAs lie in the dust-`forbidden' zone of Prantzos \& Boissier 
(\cite{prantzos00}) where 
${\rm [Zn/H]} + \log N{\rm (\HI)} > 21$
(see Figure 5).

(2) The dust-to-metals ratio, as measured by the quantity
[Cr/Zn], exhibits no systematic difference between the two
samples---we unearthed no evidence to show that radio-selected 
QSOs should be more reddened by intervening systems than
optically selected QSOs.

These results, together with recent reports of low dust 
extinction in large DLA samples drawn from the Sloan 
Digital Sky Survey (e.g. Murphy \& Liske 2004), make it 
increasingly difficult to appeal to dust-induced selection 
effects to explain the observed properties of DLAs.
It seems unlikely, for example, that the true metallicity 
of DLAs may have been underestimated by as much as a factor 
of five, as recently claimed by Vladilo \& P\'{e}roux (2005).
Below the redshift limit of our sample, $z < 1.8$, the 
situation is less clear.  The dust fraction may be higher 
(e.g. Vladilo 2004), but the number density of \MgII\ 
absorbers from an extension of the CORALS radio-selected 
sample (in the range $0.6 < z < 1.7$) is in excellent agreement 
with that from optically selected surveys (Ellison 2005), 
suggesting that dust obscuration is not a problem at these lower 
redshifts either.

One caveat (often mentioned when considering the
results of the CORALS survey) is the limited size of the 
CORALS DLA sample: are the 20 DLAs in Table 5
representative of the population as a whole, 
or are we being thrown off course by a statistical
fluctuation?
Clearly, only future observations
of a larger sample of radio selected (or possibly X-ray selected) 
QSOs will settle
this issue and test, for instance, whether the higher
metallicity---by 0.21\,dex---reported here is a 
real difference, or possibly even an underestimate
of the true value of ${\rm [} \langle{\rm (Zn/H)}_{\rm DLA}\rangle {\rm ]}$
in an unbiased sample of DLAs.
For the moment, in the absence of better statistics, we 
can use a Monte-Carlo approach to address such concerns, 
particularly if we wish to test for differences at a level 
as high as the factor of five claimed by 
Vladilo \& P\'{e}roux (\cite{vladilo05}).

Specifically, we have investigated the likelihood of 
measuring a column density weighted metallicity
${\rm [} \langle{\rm (Zn/H)}_{\rm DLA}\rangle {\rm ]} \leq -0.88$,
as found here, by drawing 20 DLAs at random from 
a much larger parent sample of DLAs 
with the column density and metallicity 
distributions proposed by Vladilo \& P\'{e}roux (\cite{vladilo05}).
These authors approximated the true
\ion{H}{i} column density distribution with a 
power law of the form 
$f_{N({\rm H~I})} \propto N$(\ion{H}{i})$^{-\beta}$,
and the true metallicity distribution with a Schechter
function $f_{Z} \propto (Z/Z_{\ast})^{\alpha}\,e^{-Z/Z_{\ast}}$
(motivated by luminosity-metallicity relationship of galaxies
in the local universe).
The least extreme set of parameters among those considered by
Vladilo \& P\'eroux has 
$\beta = 1.6$, $\alpha = -0.46$ and 
$\log (Z_{\ast}/Z_{\odot}) = -0.19$;
with these values, their column density weighted 
metallicity is 
${\rm [} \langle{\rm (Zn/H)}_{\rm DLA}\rangle {\rm ]} = -0.44$.
Our simulations showed that, drawing 20 DLAs at random
from this parent population, one would find by chance values
of ${\rm [} \langle{\rm (Zn/H)}_{\rm DLA}\rangle {\rm ]}$
as low as $-0.88$, or lower, only in five cases out of a hundred.

The absence of a detectable dust-related bias in current
magnitude-limited samples of optically selected QSOs
may appear surprising to some. After all, such a bias
has long been advocated by theorists to improve the
match of their models to the properties of
DLAs (e.g. Cen et al. \cite{cen03};
Churches et al. \cite{churches04}; 
Nagamine et al. \cite{nagamine04}).
It also seemed the 
natural explanation for the
empirical lack of DLAs with high column densities of metals
highlighted by Boiss\'{e} et al. (\cite{boisse98}) 
and still present in the larger sample 
assembled by Kulkarni et al. (2005) and indeed in
the CORALS sample considered here (see Figure 5).
The reason why most optical QSO samples do not underestimate
significantly quantities such as $\Omega_{\rm DLA}$ and 
$Z_{\rm DLA}$ compared with CORALS
was clarified by Ellison et al. (\cite{ellison04})
and is related to the shape of the QSO
luminosity function.
Optical QSO surveys will not yield significantly
skewed DLA statistics provided they reach below
the break in the QSO luminosity function at $B \sim 19$.
Brighter QSO samples, on the other hand, may show a bias
from either dust extinction or lensing (the two effects would
of course operate in different directions).

The main conclusion of the CORALS project so far---that
there are only minor differences, if any, between DLA
samples drawn from QSOs surveys at radio and at optical
wavelengths---can only be regarded as `good news'.
Its corollary is that the large data samples being made
available by major projects such as the Two-degree Field
and the Sloan Digital Sky Survey afford us an unfettered
view of the absorber population, although surveys which
concentrate only on bright QSOs may not.
The challenge is now to understand, perhaps with more
focused theoretical efforts, the rightful place of
damped Ly$\alpha$ systems within the diverse
population of galaxies known to inhabit the
high redshift universe.

\begin{acknowledgements}

We are very grateful to Varsha Kulkarni for providing us 
with her compilation of Zn and Cr measurements, 
to Kurt Adelberger and Naveen Reddy for their assistance with 
the Keck observations, and to the anonymous referee whose
suggestions improved the paper. We also acknowledge 
useful discussions with Giovanni Vladilo.
We wish to recognize 
the significant cultural role and reverence that the 
summit of Mauna Kea has always 
had within the indigenous Hawaiian community.  
We are most fortunate to have the opportunity 
to conduct observations from this mountain.

\end{acknowledgements}


{}


\begin{thebibliography}{}


\bibitem[2005]{akerman05}
Akerman, C.~J. 2005, in preparation

\bibitem[2004]{asplund04}
Asplund, M., Grevesse, N., \&\ Sauval, J. 2004, 
in Cosmic abundances as records of stellar evolution and nucleosynthesis, 
eds. F.~N. Bash \&\ T.~G. Barnes, ASP conf. series, in press (astro-ph/0410214)

\bibitem[1993]{bergeson93}
Bergeson, S.~D. \&\ Lawler J.~E. 1993, ApJ, 408, 382

\bibitem[2003]{bernstein03}
Bernstein, R., Shectman, S.~A., Gunnels, S.~M., Mochnacki, S., \&\ Athey, A.~E. 2003, SPIE, 4841, 1694

\bibitem[1998]{boisse98} 
Boiss\'{e}, P., Le Brun, V., 
Bergeron, J., \& Deharveng, J.\ 1998, \aap, 333, 841 

\bibitem[2003]{boissier03} 
Boissier, S., P{\' e}roux, C., \& Pettini, M.\ 2003, \mnras, 338, 131 

\bibitem[2004]{bresolin04} 
Bresolin, F., Garnett, D.~R., \& Kennicutt, R.~C.\ 2004, \apj, 615, 228 

\bibitem[2004]{cayrel04} 
Cayrel, R., et al.\ 2004, \aap, 416, 1117 

\bibitem[2003]{cen03}
Cen, R., Ostriker, J.~P., Prochaska, J.~X., \&\ Wolfe, A.~M. 2003, ApJ, 598, 741

\bibitem[2003]{centurion03}
Centuri{\' o}n, M., Molaro, P., Vladilo, G., P{\' e}roux, C., Levshakov, S. A., \& D'Odorico, V. 2003, A\&A, 403, 55

\bibitem[2003]{chen03} 
Chen, H., \& Lanzetta, K.~M.\ 2003, \apj, 597, 706 

\bibitem[2005]{chen05} 
Chen, H., Kennicutt, R.~C., \& Rauch, M. 2005, ApJ, 620, 703

\bibitem[2004]{chen04}
Chen, Y.~Q., Nissen, P.~E., \& Zhao, G. 2004, A\&A, 425, 697

\bibitem[2005]{christensen05} 
Christensen, L., Schulte-Ladbeck, R.~E., S{\' a}nchez, S.~F., Becker, T., 
Jahnke, K., Kelz, A., Roth, M.~M., \& Wisotzki, L.\ 2005, \aap, 429, 477 

\bibitem[2004]{churches04}
Churches, D.~K., Nelson, A.~H., \&\ Edmunds, M.~G. 2004, MNRAS, 347, 1234

\bibitem[2000]{dekker00}
Dekker, H., D'Odorico, S., Kaufer, A., Delabre, B., \& Kotzlowski, H. 2000, in SPIE Proc. 4008, 534

\bibitem[2004]{de_mello04} 
de Mello, D.~F., 
Daddi, E., Renzini, A., Cimatti, A., di Serego Alighieri, S., Pozzetti, L., 
\& Zamorani, G.\ 2004, \apjl, 608, L29 

\bibitem[1993]{efron93}
Efron, B. \& Tibshirani, R.~J. 1993, An Introduction to the Bootstrap (New York: Chapman \&\ Hall)

\bibitem[2005]{ellison2005}
Ellison, S.~L. 2005, Proceedings of IAUC 199: Probing Galaxies through Quasar Absorption Lines, 
eds. P.~R. Williams, C. Shu, and B. Menard, in press (astro-ph/0505111)

\bibitem[2004]{ellison04}
Ellison, S.~L., Churchill, C.~W., Rix, S.~A., \&\ Pettini, M. 2004, ApJ, 615, 118

\bibitem[2005a]{ellison05a}
Ellison, S.~L., Hall, P.~B., \& Lira, P. 2005a, AJ, submitted

\bibitem[2005b]{ellison05} 
Ellison, S.~L., Kewley, L.~J., \& Mall\'{e}n-Ornelas, G.
2005b, \mnras, 357, 354

\bibitem[2001a]{ellison01a}
Ellison, S.~L. \& Lopez, S. 2001a, A\&A, 380, 117

\bibitem[2001b]{ellison01}
Ellison, S.~L., Yan, L., Hook, I.~M., Pettini, M., Wall, J.~V., \&\ Shaver, P. 2001b, A\&A, 379, 292

\bibitem[1993]{fall93}
Fall, S.~M. \&\ Pei, Y.~C. 1993, ApJ, 402, 479

\bibitem[1985]{feigelson85}
Feigelson, E.~D., \&\ Nelson, P.~I. 1985, ApJ, 293, 192

\bibitem[1988]{heisler88} 
Heisler, J., \& Ostriker, J.~P.\ 1988, \apj, 332, 543 

\bibitem[1974]{hobbs74} 
Hobbs, L.~M.\ 1974, \apj, 191, 381 

\bibitem[2003]{hook03} 
Hook, I.~M., Shaver, P.~A., Jackson, C.~A., Wall, J.~V., 
\& Kellermann, K.~I.\ 2003, \aap, 399, 469 

\bibitem[2002]{jackson02} 
Jackson, C.~A., Wall, J.~V., Shaver, P.~A., Kellermann, K.~I., Hook, 
I.~M., \& Hawkins, M.~R.~S.\ 2002, \aap, 386, 97 

\bibitem[2004]{khare04}
Khare, P., Kulkarni, V.~P., Lauroesch, J.~T., York, D.~G., Crotts, A.~P.~S., \&\ Nakamura, O. 
2004, ApJ, 616, 86

\bibitem[2005]{kulkarni05}
Kulkarni, V.~P., Fall, S.~M., Lauroesch, J.~T., York, D.~G., Welty, D.~E., Khare, P., \&\ Truran, J.~W. 
2005, ApJ, 618, 68

\bibitem[1981]{latta81}
Latta, R.~B. 1981, J. Am. Statistical Association, 26, 713

\bibitem[1992]{lavalley92}
LaValley, M., Isobe, T., \&\ Feigelson, E.~D. 1992, ``ASURV'', BAAS, 24, 839

\bibitem[2003]{lodders03}
Lodders, K. 2003, ApJ, 591, 1220

\bibitem[1996]{lu96}
Lu, L., Sargent, W.~L.~W., Barlow, T.~A., Churchill, C.~W., \& Vogt, S.~S. 1996, ApJS, 107, 475

\bibitem[1989]{meyer89}
Meyer, D.~M., Welty, D.~E., \&\ York, D.~G. 1989, ApJ, 343, L37

\bibitem[1998]{mo98}
Mo, H.~J., Mao, S., \&\ White, S.~D.~M. 1998, MNRAS, 295, 319

\bibitem[2003]{morton03}
Morton, D.~C. 2003, ApJS, 149, 205

\bibitem[2004]{murphy04}
Murphy, M.~T. \&\ Liske, J. 2004, MNRAS, 354, L31

\bibitem[2004]{nagamine04} 
Nagamine, K., Springel, V., \& Hernquist, L.\ 2004, \mnras, 348, 435 

\bibitem[1995]{oke95}
Oke, J.~B., Cohen, J.~G., Carr, M., Cromer, J., Dingizian, A., \&\ Harris, F.~H. 1995, PASP, 107, 375 

\bibitem[1995]{pei95} 
Pei, Y.~C., \& Fall, S.~M.\ 1995, \apj, 454, 69 
\bibitem[1991]{pei91}
Pei, Y.~C., Fall, S.~M., \&\ Bechtold, J. 1991, ApJ, 378, 6

\bibitem[2003]{peroux03}
P{\'e}roux, C., McMahon, R.~G., Storrie-Lombardi, L.~J., \& Irwin, M.~J. 2003, MNRAS, 346, 1103

\bibitem[2004]{pettini04} 
Pettini, M.\ 2004, in Cosmochemistry.~The melting pot of the elements, 
eds. C.~Esteban, R.J. Garc\'{\i}a L\'{o}pez, A. Herrero, 
\& F. S\'{a}nchez,
(Cambridge, Cambridge University Press), 257 
\bibitem[1989]{pettini89}
Pettini, M., Boksenberg, A., \&\ Hunstead, R.~W. 1989, NATO ASIC Proc.~264: The Epoch of 
Galaxy Formation, 107  
\bibitem[1990]{pettini90}
Pettini, M., Boksenberg, A., \&\ Hunstead, R.~W. 1990, ApJ, 348, 48
\bibitem[1999]{pettini99}
Pettini, M., Ellison, S.~L., Steidel, C.~C., \&\ Bowen, D.V. 1999, ApJ, 510, 576
\bibitem[1997a]{pettini97a}
Pettini, M., King, D.~L., Smith, L.~J.,  \&\ Hunstead, R.~W. 1997a, ApJ, 478, 536
\bibitem[2002]{pettini02} 
Pettini, M., Rix, S.~A., Steidel, C.~C., Adelberger, K.~L., Hunt, M.~P., \& Shapley, A.~E.\ 
2002, \apj, 569, 742 
\bibitem[1994a]{pettini94}
Pettini, M., Smith, L.~J., Hunstead, R.~W., \&\  King, D.~L. 1994, ApJ, 426, 79
\bibitem[1997b]{pettini97b}
Pettini, M., Smith, L.~J., King, D.~L., \&\ Hunstead, R.~W. 1997b, ApJ, 486, 665

\bibitem[2000]{prantzos00} 
Prantzos, N., \& Boissier, S.\ 2000, \mnras, 315, 82 

\bibitem[2003]{prochaska03a} 
Prochaska, J.~X. 2003, ApJ, 582, 49 
\bibitem[2001]{prochaska01} 
Prochaska, J.~X., et al.\ 2001, \apjs, 137, 21 
\bibitem[2003]{prochaska03} 
Prochaska, J.~X., Gawiser, E., Wolfe, A.~M., Castro, S., 
\&\ Djorgovski, S.~G.\ 2003, \apjl, 595, L9 
\bibitem[1999]{prochaska99}
Prochaska, J.~X., \&\ Wolfe, A.~M. 1999, ApJS, 121, 369
\bibitem[2000]{prochaska00}
Prochaska, J.~X., \&\ Wolfe, A.~M. 2000, ApJ, 533, L5
\bibitem[2002]{prochaska02} Prochaska, J.~X., 
\& Wolfe, A.~M.\ 2002, \apj, 566, 68 

\bibitem[2003]{rao03} 
Rao, S.~M., Nestor, D.~B., Turnshek, D.~A., Lane, W.~M., Monier, E.~M., 
\& Bergeron, J.\ 2003, \apj, 595, 94 
\bibitem[2000]{rao00}
Rao, S.~M. \&\ Turnshek, D.~A. 2000, ApJS, 130, 1

\bibitem[2003]{ryan-weber03}
Ryan-Weber, E.~V., Webster, R.~L., \&\ Staveley-Smith, L. 2003, MNRAS, 343, 1195

\bibitem[1996]{savage96} 
Savage, B.~D., \& Sembach, K.~R.\ 1996, \araa, 34, 279 

\bibitem[2002]{sheinis02}
Sheinis, A.~I., Bolte, M., Epps, H.~W., Kibrick, R.~I., Miller, J.~S., Radovan, M.~V., Bigelow, B.~C., \&\ Sutin, B.~M. 2002, PASP, 114, 851

\bibitem[2004]{shapley04} 
Shapley, A.~E., Erb, 
D.~K., Pettini, M., Steidel, C.~C., \& Adelberger, K.~L.\ 2004, \apj, 612, 108 

\bibitem[2000]{storrie00}
Storrie-Lombardi, L.~J., \& Wolfe, A.~M. 2000, ApJ, 543, 552

\bibitem[2002]{stoughton02}
Stoughton, C. et al. 2002, AJ, 123, 485

\bibitem[2005]{swinbank05} 
Swinbank, A.~M., Smail, I., Chapman, S.~C., Blain, A.~W., 
Ivison, R.~J., \& Keel, W.~C. 2005, ApJ, 617, 64

\bibitem[2004]{vladilo04} 
Vladilo, G.\ 2004, \aap, 421, 479 
\bibitem[2005]{vladilo05} 
Vladilo, G., \& P\'{e}roux, C. \ 2005, \aap, 
submitted (astro-ph/0502137) 

\bibitem[2005]{weatherley05} 
Weatherley, S.~J., Warren, S.~J., M$\o$ller, P., Fall, S.~M., 
Fynbo, J.~U., \& Croom, S.M.  2005, MNRAS, in press (astro-ph/0501422)

\bibitem[1995]{wolfe95}
Wolfe, A.~M., Lanzetta, K.~M., Foltz, C.~B., \&\ Chaffee, F.~H. 1995, ApJ, 454, 698
\bibitem[1986]{wolfe86}
Wolfe, A.~M., Turnshek, D.~A., Smith, H.~E., \& Cohen, R.~D. 1986, ApJS, 61, 249

\end{thebibliography}
\end{document}